\documentclass[pdflatex,sn-mathphys-num]{sn-jnl}
\usepackage{graphicx}
\usepackage{multirow}%
\usepackage{amsmath,amssymb,amsfonts}%
\usepackage{amsthm}%
\usepackage{mathrsfs}%
\usepackage[title]{appendix}%
\usepackage{xcolor}%
\usepackage{textcomp}%
\usepackage{manyfoot}%
\usepackage{booktabs}%
\usepackage{algorithm}%
\usepackage{algorithmicx}%
\usepackage{algpseudocode}%
\usepackage{listings}%
\usepackage{subfigure}
\usepackage{rotating}
\usepackage{pdflscape}
\usepackage{longtable}
\usepackage{makecell}
\usepackage{pdfpages}

\theoremstyle{definition}

\theoremstyle{plain}
\newtheorem{lemma}[]{Lemma}

\begin{document}

\title[Article Title]{Parametrization of Microbial Survival Models under
UVC Exposure}

\author[1]{\fnm{Aikaterini~A.} \sur{Tsantari}}\email{ktsantari@auth.gr}

\author*[2]{\fnm{Konstantinos~K.} \sur{Delibasis}}\email{kdelimpasis@uth.gr}
\equalcont{These authors contributed equally to this work.}

\author[2]{\fnm{Harilaos~G.} \sur{Sandalidis}}\email{sandalidis@uth.gr}
\equalcont{These authors contributed equally to this work.}

\author[1]{\fnm{Nestor~D.} \sur{Chatzidiamantis}}\email{nestoras@auth.gr}
\equalcont{These authors contributed equally to this work.}

\affil[1]{\orgdiv{Department of Electrical and Computer Engineering}, \orgname{Aristotle University of Thessaloniki}, \orgaddress{\city{Thessaloniki}, \postcode{54124}, \state{Macedonia}, \country{Greece}}}

\affil*[2]{\orgdiv{Department of Computer Science and Biomedical Informatics}, \orgname{University of Thessaly}, \orgaddress{\street{Papasiopoulou 2-4}, \city{Lamia}, \postcode{35131}, \state{Central Greece}, \country{Greece}}}

\abstract{
This work presents a unified framework for the parametrization and comparison of microbial survival models under UVC exposure. Four dose–response models—the single-target, multi-target, linear–quadratic, and two-stage decay models—are analyzed with respect to their mathematical structure, parameter identifiability, and biological interpretability, while ensuring physically meaningful parameter estimates. To address the limited size and lack of replication in published datasets, parameter uncertainty is quantified using a parametric bootstrap approach under multiplicative dose uncertainty. Model comparison combines goodness-of-fit in logarithmic survival space, the Akaike Information Criterion corrected for small samples, and identifiability considerations. Application to data for 32 microorganisms shows that no single model is universally optimal, highlighting that reliable model selection requires combining statistical performance with physical and biological consistency, and providing a robust basis for UVC survival analysis and disinfection modeling. }
\keywords{microorganisms, parametrization, survival models, UVC exposure}

\maketitle

\section{Introduction}
Ultraviolet (UV) radiation covers the wavelength range of 100 to 400 nm and is divided into three  subcategories: UVA (315-400 nm), UVB (280-315 nm), and UVC (100-280 nm) \cite{Vavoulas2019}. UVC, in particular, is harmful to humans, causing skin and eye damage as well as cancer \cite{Turner2018}. Specific wavelengths within the UV spectrum can sterilize air and surfaces. This application, known as ultraviolet germicidal irradiation (UVGI), primarily utilizes UVB and UVC wavelengths (200-320 nm) \cite{Kowalski2010}.

UVC radiation is absorbed by proteins, RNA, and DNA, resulting in cellular damage. This absorption induces mutations, often leading to the inactivation or death of microorganisms. Microbial inactivation refers to losing the ability to reproduce or multiply without necessarily damaging its genetic material, rendering the organism non-pathogenic \cite{Lelieveld2020}. During partial dormancy, the microbe can reactivate, initiating multiplication after UVC exposure, utilizing various cellular mechanisms \cite{Demeersseman2023}. Microbial death occurs upon receiving a sufficient dose of radiation, termed the lethal dose. Although the inactivation mechanism applies to all microorganisms, including fungi, bacteria, or viruses, their sensitivity to UV exposure may vary significantly. Consequently, the lethal dose or exposure time can vary between microorganisms \cite{Wang2009}. Hence, the effectiveness of UVC radiation in disinfecting a microorganism depends not only on the exposure time but also on the specific genomic regions affected \cite{Raeiszadeh2020}.

The survival of microorganisms on surfaces is influenced by various factors, including the types, structures, and chemical compositions of the microorganisms. Consequently, microorganisms within the same family, such as Sars-CoV-1 and Sars-CoV-2, exhibit similar lethal doses \cite{Raeiszadeh2020}, \cite{Arguelles2020}. Environmental conditions such as temperature, relative humidity, and surface material also play significant roles \cite{Bhardwaj2020,Vasickova2010,Lindblad2020}, with relative humidity having different effects on viruses and bacteria, while its influence on fungi is minimal \cite{Kowalski2010}. The characteristics of the surface, including its color, texture, and density, are crucial in determining the absorption or reflection of photons, thereby affecting the amount of radiation it absorbs \cite{Turner2018}, \cite{Lindblad2020}. Moreover, the correct arrangement of light sources in a room is paramount in practical settings \cite{Botta2020}. Finally, it is critical to recognize that microorganism distribution within a space is dynamic and subject to variation \cite{Thatcher2022}.

For some microorganisms, different doses of UV radiation were found to cause different reductions in their cell colonies. A survival model is a mathematical expression of survival versus radiation dose. Given a parameterized survival model for a microorganism, it is possible to calculate its survival fraction (equivalently, the reduction) for any radiation dose within a specific range. The curve for this model is usually plotted on a logarithmic scale for the y-axis. The shape of the curve varies depending on the mathematical formula chosen to express it \cite{Kowalski2010}.

Overall, survival models for microorganisms under UVC radiation are indispensable tools for enhancing the safety, efficiency, and sustainability of disinfection methods. In research, they help understand UV damage mechanisms and foster innovation in UVC technology. Additionally, they aid in designing safety protocols for UVC exposure and devising decontamination strategies in emergencies, making them essential across various fields and applications. 

Inspired by these considerations, the present study develops a unified mathematical and statistical framework for the analysis of microbial survival under UVC irradiation. Specifically, we examine four widely used dose--response models, namely the Single-Target (ST), Multi-Target (MT), Linear--Quadratic (LQ), and Two-Stage Decay (TSD) models. Rather than focusing solely on curve fitting, the study investigates the mathematical properties of these models, the biological interpretability of their parameters, and the conditions under which their parameterizations remain physically meaningful and identifiable.

Using published UVC dose--survival data for 32 microorganisms, we estimate the parameters of all four models and quantify their uncertainty through a parametric bootstrap approach. Model comparison is performed using complementary criteria, including the mean squared error in log-survival space, the Akaike Information Criterion (AIC) corrected for small sample sizes (AICc), parameter identifiability, and biological plausibility. The aim of this work is, therefore, not to derive universal, condition-independent parameters, but to provide a consistent framework for comparing survival models, selecting the most appropriate one for a given microorganism and dataset, and supporting future UVC dose--response modeling studies.

\section{UVC disinfection studies}

Various investigations explore the inactivation of microorganisms using UVC radiation. Some studies involve practical measurements in real-world settings, such as \cite{Lindblad2020}, \cite{Botta2020}, and \cite{Cao2020,Baluja2020,Ngom2022}, while others focus on experimental measurements conducted in laboratory conditions, for instance, \cite{Wang2009} and \cite{Narita2020,Buonanno2020,Chiappa2021,Hessling2020}. All these studies assess the survival of microorganisms, quantified on a logarithmic scale based on 10. For example, a 1log reduction corresponds to a 90\% reduction (0.1 survival) from the initial microbial population \cite{LogReduction2024}. Measurement involves counting the colony-forming units (CFU) initially present and those remaining after exposure. 

\begin{sidewaystable*}[p]
\centering
\caption{Summary of irradiation conditions for the studies used in model fitting}
\label{tab:uv_conditions}
\footnotesize
\setlength{\tabcolsep}{4pt}
\renewcommand{\arraystretch}{1.12}

\resizebox{\textheight}{!}{
\begin{tabular}{p{3.7cm} p{1.5cm} p{1.7cm} p{3.0cm} p{4.2cm} p{2.2cm} p{3.6cm} p{1.8cm}}
\toprule \toprule
\textbf{Type} &
\textbf{Application} &
\textbf{Wavelength} &
\textbf{UV Source} &
\textbf{Apparatus and Sample Holding} &
\textbf{Distance} &
\textbf{Sensor / Radiometer} &
\textbf{Exp. Time} \\
\midrule \midrule

% Beta 
\cite{liltved1996influence} \textit{Yersinia ruckeri} &
aquaculture safety &
254 monochr &
15-W low-pressure (LP) mercury lamp (Philips) &
collimated beam apparatus; bacterial suspensions (50 mL) were placed in Petri dishes and slowly stirred &
20--30 cm &
calibrated UVX-25 sensor coupled to a UVX radiometer (UVP Inc.) &
2--30 s \\
\midrule 
% gamma
\makecell[l]{\cite{lindblad2020ultraviolet} \textit{C. jejuni ATCC 43429},\\ \textit{E. coli O157.H7 ATCC 43894},\\ textit{S. typhi ATCC 19430},\\ \textit{S dysenteriae ATCC29027},\\ \textit{L. pneumophila ATCC 43660}} &
Hospital infections &
254 monochr &
Low-pressure (LP) mercury vapor germicidal lamps &
collimated beam apparatus; purified and suspended in deionized water &
-- &
International Light IL-700 radiometer / SEE-240 detector &
30 s -- min \\
\midrule
% D
\cite{giese2000sensitivity} \textit{Citrobacter diversus} \textit{Citrobacter freundii}, \textit{Klebsiella pneumoniae} &
Water safety &
254, 280, 301 &
medium-pressure (MP) mercury vapor lamp (1 kW), with monochromator &
collimated beam apparatus; microorganisms were suspended in a phosphate-buffered solution (pH 7.0) and stirred &
20--40 cm &
UV-visible spectrophotometer &
sec to min \\
\midrule
% E 
\textit{Escherichia coli O157.H7 CCUG 29193}, \textit{CCUG 29197}, \textit{CCUG 29199}. \textit{ATCC 25922}, \textit{K12 IFO3301} \cite{sommer1998time} &
Water safety &
253.7 &
10 low-pressure mercury lamps, regulated intensity &
collimated beam apparatus; cells suspended in sterile 0.85\% saline; aliquots of 25 mL were irradiated in Petri dishes &
30 cm &
-- &
2--102 s \\

\midrule
% THETA 
\textit{Legionella pneumophila ATCC33152A},\textit{B}, \cite{oguma2002photoreactivation} &
Water safety &
200--300 polychromatic &
medium-pressure (MP) mercury vapor lamp &
cells suspended in phosphate buffer; 40 mL of the suspension in a 100 mm Petri dish &
-- &
multichannel photodetector (MCPD-2000; Otsuka) &
0--30 s \\
\midrule
% Z
\textit{Escherichia coli ATCC 11229}, \cite{hoyer1998testing} &
water safety &
254 &
Low-pressure mercury lamps &
collimated beam apparatus; suspended in water with varying UV transmittance (UVT), passing through UVC reactor &
bench scale: 20--40 cm; full-scale reactor: most unfavorable point &
calibrated UV-C radiometer with a silicon carbide (SiC) or gallium nitride (GaN) photodiode &
1--5 s \\
\midrule
% H
\textit{Escherichia coli ATCC 11303}, \cite{wu2005impacts} &
water safety &
254 &
15 WLow-pressure mercury lamp &
Goethite (a-FeooH) -cell suspensions, then resuspended in phosphate-buffered saline and incubated for 24 h at 20°C.
Put on shallow Petri dish &
-- &
-- &
-- \\
\midrule
% IOTA 
\textit{Escherichia coli O157.H7}, \cite{yaun2003response} &
food safety &
254 monochr &
single G36T6 Model 4136 germicidal light unit &
UV chamber 1 m long; plates containing tryptic soy agar plus 50 ppm nalidixic acid (TSAN) &
1 m &
dosimeter calibrated specifically at 253.7 nm (Spectronics, Westbury, N.Y.) &
5--75 s \\
\hline
% Lambda 
\textit{Pseudomonas stutzeri}, \textit{Vibrio natriegens} \textit{Salmonella typhimurium}, \cite{joux1999marine} &
Marine Bacteria &
UVB 280--320 &
UVB-enhanced fluorescent lamp &
bacteria suspended in artificial seawater, quartz tubes or open dishes &
-- &
SED240 solar-blind vacuum photodiode detector &
up to 6 h \\
\midrule
% M 
\textit{Salmonella spp.}, \textit{Salmonella typhimurium hf}, \cite{tosa1999photoreactivation} &
-- &
254 mono &
low-pressure (LP) germicidal lamp &
collimated beam apparatus; microorganism in phosphate-buffered saline (PBS) &
-- &
calibrated UVC radiometer for 254 nm &
-- \\
\midrule
% N 
\textit{Shigella sonnei ATCC9290}, \textit{Staphylococcus aureus ATCC25923}, \textit{Streptococcus faecalis ATCC29212}, \textit{Salmonella typhi ATCC 6539} \cite{chang1985uv} &
Epidemiology &
254 nm (monochromatic) &
UVB-enhanced fluorescent lamps (UVB-313; Q-Panel Co.) &
collimated beam apparatus; microbes suspended in a phosphate buffer, stirrable &
-- &
International Light IL-500 radiometer / SEE-240 detector &
few seconds up to several minutes \\
\midrule
% KSI 
\textit{Streptococcus faecalis se}, \cite{harris1987influence} &
water safety &
254 mono &
low-pressure (LP) mercury vapor germicidal lamps &
batch-scale UV reactor for secondary wastewater effluent &
-- &
International Light IL-700 radiometer / SEE-240 detector &
1 and 30 s \\
\hline
% X1
\textit{Bacillus subtillis ATCC6633}, \cite{Lindblad2020}, \cite{marshall2003comparison} &
water safety &
254 mono &
low-pressure (LP) mercury lamps &
collimated beam apparatus; spores suspended in deionized water or PBS and placed on rotating 6-cm Petri dish&
-- &
International Light IL1700 Research Radiometer; detector: SED240 &
1--5 s \\
\hline\

% \textit{Bacillus subtilis ATCC6633} ??
X2 &
water safety &
254 mono &
low-pressure (LP) mercury lamps &
collimated beam apparatus; microbes suspended in deionized water &
-- &
-- &
30--120 s \\
\midrule
% Salmonella enterica subsp. Enterica serovar Typhimurium
\cite{narita2020ultraviolet} 
\textit{Salmonella ent. sbs. serovar Typhimurium} &
Epidemics &
254 &
a 3mW/cm2 254 nm Lamp &
Suspension on tryptic soy agar, placed in wells of a 24-well plate placed on ice &
6cm &
-- &
time to achieve 0.01 -- 0.8J/cm2 \\
\midrule
% SARS-COV 2
\cite{Raeiszadeh2020} \textit{SARS CoV-2} &
Epidemics &
254 mono &
CL-1000 UVP Crosslinker consisted of 5  ×  254  nm shortwave tubes &
Microbiological safety cabinet.
Virus stocks in wells of a 24-well plate placed on ice &
6cm &
-- &
time to achieve 0.01 -- 0.8J/cm$^2$ \\

\bottomrule \bottomrule
\end{tabular}
}
\label{Table1}
\end{sidewaystable*}

The most prevalent wavelengths of UV radiation for microorganism inactivation are 222 nm and 254 nm. Recent studies \cite{Narita2020}, \cite{Buonanno2020} suggest that 222 nm radiation is less harmful to humans because it penetrates only superficially and does not breach the cell walls of skin cells or the outer layer of the eye. Nevertheless, it retains its sterilizing efficacy, making it advisable for use in public places \cite{Buonanno2020}. Conversely, wavelengths less than 240 nm (in the far-UVC band) generate ozone through the interaction of photons with oxygen molecules, leading to harmful effects and significant air pollution \cite{Turner2018}, \cite{Demeersseman2023}, \cite{Raeiszadeh2020}.

As depicted in \cite{Raeiszadeh2020}, coronaviruses exhibit higher UV radiation resistance than \textit{Escherichia coli} bacteria, necessitating significantly greater radiation doses for inactivation. Moreover, \cite{Narita2020} asserts that 222 nm is particularly effective against bacterial endospores, whereas 254 nm yields superior results for fungal spores and hyphae.

Details of UVC irradiation experiments have been documented in the literature, based on laboratory measurements or experimental assessments conducted in real-world settings \cite{Raeiszadeh2020}, \cite{Lindblad2020}, \cite{Baluja2020}, \cite{Chiappa2021}. Specifically, \cite{Lindblad2020} and \cite{Baluja2020} provide measurements for various common microorganisms, including \textit{Bacillus subtilis} \textit{ATCC6633}, \textit{Campylobacter jejuni ATCC43429},\textit{ Citrobacter diversus}, \textit{Escherichia coli}, \textit{Klebsiella pneumoniae}, \textit{Legionella pneumophila ATCC33152}, \textit{Shigella dysenteriae}, \textit{Shigella sonnei ATCC9290}, \textit{Staphylococcus aureus ATCC25923}, and\textit{ Streptococcus faecalis ATCC29212}. Furthermore, data regarding air disinfection of microorganisms and measurements involving animal cells are available in \cite{Chiappa2021}. While air decontamination shares similarities with surface decontamination, both differ significantly from water decontamination \cite{Turner2018}, \cite{Darnell2004}.

It is notable to reference the findings of \cite{Baluja2020}, which explore the disinfection of face masks using the DIALux simulator for theoretical sterilization studies. The research demonstrates the time required to disinfect masks with specific radiation doses and discusses that the region nearest to the source does not always receive the highest dose and, consequently, the most effective disinfection. 

In this work, iraadiation and survival data for several microorganisms were compiled from multiple published studies. 
The selection of microorganisms was primarily driven by the availability of sufficiently detailed dose–response data with measurements for a number of doses in the published literature, which is necessary for reliable fitting of multi-parameter survival models. Most of the selected microorganisms are important in the field of epidemiology, food and water safety.
Table~\ref{Table1} summarizes UVC irradiation conditions, equipment and other experimental details, along with the relevant literature references, for all microorganisms studied. In cases where the original studies did not report key parameters such as distance from the source or exposure time, these are indicated with a dash. 

\section{Survival models}

Mathematical modeling of microorganism survival often assumes that each microorganism has one or more targets that must be hit for its complete inactivation, leading to the development of corresponding models such as the ST, MT, and LQ. A target is a sensitive region in the DNA molecule whose exposure to UV radiation can cause cell death \cite{CellSurvivalCurves2024}. The simplest survival model is the ST model with one target. Its mathematical expression is given by

\begin{equation}
S = e^{-D/D_0},
\label{Eq_ST}
\end{equation}
where $D$ is the dose variable, and $D_0$ is the dose required to reduce cell survival by a factor of $e^{-1}$. Radiative emissions are assumed to be Poisson-distributed~\cite{McMahon2018}. Survival curves are commonly plotted on a semi-logarithmic graph. Substituting $\alpha = 1/D_0 < 0$, (1) transforms to
\begin{equation}
S = e^{-a_{ST} D},
\label{Eq_ST-model}
\end{equation}
where the coefficient $a$ takes different values for each microorganism. The MT model~\cite{Kowalski2010}, \cite{CellSurvivalCurves2024}, \cite{Nomiya2013} is given by the mathematical expression
\begin{equation}
S = 1 - \left(1 - e^{-D/D_0}\right)^{b_{MT}},
\end{equation}
where $b_{MT}$ represents the number of targets per cell that must be hit to deactivate the cell~\cite{CellSurvivalCurves2024}, \cite{Nomiya2013}. This model is particularly applicable to microorganisms exposed to low linear energy transfer (LET) radiation, such as UVC photons, depending on the value of $\beta$~\cite{CellSurvivalCurves2024}. In this case, where $b_{MT} > 1$, a shoulder appears in the region of low doses in the survival curve~\cite{McMahon2018}. For this model, the coefficient $a_{MT}$ is again defined as the quotient $-1/D_0$. Therefore, (3) is written as
\begin{equation}
S = 1 - \left(1 - e^{a_{MT} D}\right)^{b_{MT}}.
\label{Eq_MT-model}
\end{equation}
The next model analyzed in this work is the LQ, which is expressed as
\begin{equation}
S = e^{-a_{LQ} D - b_{LQ} D^2}.
\label{Eq_LQ-model}
\end{equation}
This curve also exhibits a non-linearity, which is affected by the ratio $a_{LQ}/b_{LQ}$ of the coefficients. Thus, the LQ model could be considered a combination of the previous two. According to~\cite{McMahon2018}, the LQ model often has a somewhat similar form to the MT model--especially for a few experimental points, which is also confirmed in the curves below for various microorganisms.

Experimental evidence suggests that certain microorganisms exhibit bi-phasic survival behavior: a large fraction of the population is highly susceptible to radiation, while a smaller fraction is more resistant. This leads to a rapid initial decline in population at low doses, followed by a slower decrease at higher doses. In such cases, the aforementioned models fail to adequately capture the observed behavior. A more appropriate representation is provided by the TSD, or bi-exponential model~\cite{Kowalski2010}
\begin{equation}
S = w e^{-a_{TSD} D} + (1-w) e^{-b_{TSD} D},
\label{eq_TSDmodel}
\end{equation}
where $a_{TSD}$ and $b_{TSD}$ are the inactivation rate constants (reciprocals of $D_0$ in the other models), and $w \in [0,1]$ denotes the fraction of the more sensitive subpopulation. The condition $a_{TSD} > b_{TSD} > 0$ ensures that the first term corresponds to the faster decay component.

For survival curves characterized by an initially steep and subsequently shallower $\ln S$--dose response, we introduce the crossover dose $D_\times$, defined as the point at which dominance in the instantaneous inactivation rate shifts from the fast-decaying component to the slower one. The quantity $D_\times$ provides a more direct and biologically meaningful characterization of this transition.

\begin{lemma}
    The crossover dose $D_x$ where the slopes of the two components are equal is given by \begin{equation}
D_\times
=
\frac{1}{a_{TSD}-b_{TSD}}
\ln\!\left(\frac{w a_{TSD}}{(1-w)b_{TSD}}\right).
\label{eq:Dx_slope}
\end{equation}
\end{lemma}

\begin{proof}
The local slope magnitude of the $\ln S$ versus dose curve is defined as
\begin{equation}
m(D)
=
-\frac{d}{dD}\ln S
=
\frac{w a_{TSD} e^{-a_{TSD} D}+(1-w)b_{TSD} e^{-b_{TSD} D}}
     {w e^{-a_{TSD} D}+(1-w)e^{-b_{TSD} D}}.
\label{eq:slope}
\end{equation}
This quantity decreases smoothly from
\[
m(0)=w a_{TSD}+(1-w)b_{TSD}
\]
to the asymptotic value
\[
\lim_{D\to\infty} m(D)=b_{TSD},
\]
reflecting a transition from fast to slow killing with increasing dose.    

Crossover dose is defined by equal contributions to the instantaneous killing rate, i.e., to the slope of $\ln S$
\begin{equation}
w a_1 e^{-a_{TSD} D_\times}=(1-w)b_{TSD} e^{-b_{TSD} D_\times}.
\end{equation}
This condition leads to \eqref{eq:Dx_slope}.
\end{proof}

\section{Survival model parametrization}
The survival models can be parameterized for various microorganisms using experimental measurements specific to each type~\cite{Raeiszadeh2020},~\cite{Lindblad2020}. In~\cite{Lindblad2020}, such measurements are gathered from the literature and experimentally validated in a hospital room at different distances from the source, both in areas with and without a direct line-of-sight (LOS) to the source. Similarly, a study on SARS-CoV-2 is conducted at 254 nm, one of the preferred wavelengths for sterilization~\cite{Raeiszadeh2020}.
 
In particular, for the ST model, the $i$-th experimental measurement $(D_i, S_i), i = 1, \dots, N$ of a microorganism should satisfy the model equation. By applying logarithmic transformations to these expressions and employing table-based formalism, we derive the following equation
\begin{equation}
\mathbf{D} \alpha_{ST} = \mathbf{S}_\mathbf{L},
\end{equation}
where $\mathbf{D} = [D_1, \dots, D_N]^T$ is the column matrix of the doses at the points, $\mathbf{S_L} = -[\ln S_1, \dots, \ln S_N]^T$ is the column matrix of the corresponding logarithmic survivals, and $\alpha$ is the unknown model parameter. 

Eq.~(7) is solved for $\alpha$ by linear least squares minimization, which is equivalent to multiplying from the left by $\mathbf{D}^T$. The final calculation formula for $\alpha$ is shown below
\begin{equation}
a_{ST} = -\frac{\sum_{i=1}^{N} D_i S_{L_i}}{\sum_{i=1}^{N} D_i^2},
\end{equation}
where $N$ is the total number of experimental data points for each microorganism.

\subsection{Two-step parameterization for the MT model}

To obtain stable initial values for the MT model,
$S=1-\left(1-e^{-D/D_0}\right)^{b_{MT}}=1-\left(1-e^{-a_{MT} D}\right)^{b_{MT}},$
a two-step approach was employed. First, a linear least-squares regression was performed on the deep-survival data in log space, i.e., $\ln S$ versus $D$, under the approximation that curvature effects are negligible in the low-survival regime. The slope of this regression provided an initial estimate of the characteristic dose parameter $D_0$. The intercept of the regression was then used to derive a preliminary estimate of the shoulder-related exponent parameter (via exponential transformation), yielding an initial value for $n$. First, we will prove the following Lemma.

\begin{lemma}
The gradient of the logarithm of survival for the MT model for large values of dose $D$ is independent of the number of targets per cell, $b_{MT}$.
\end{lemma}

\begin{proof}
By differentiating the logarithm of $S$ with respect to $D$ in (4), we obtain
\begin{equation}
\hspace{-9pt}\frac{d}{dD} \ln S = \frac{b(1 - e^{-D/D_0})^{b_{MT}}}{D_0 \left(e^{-D/D_0} - 1\right) \left(1 - \left(1 - e^{-D/D_0}\right)^{b_{MT}} \right)}.
\end{equation}
It is easy to show that
\begin{equation}
\mathop {\lim }\limits_{D \to \infty } \frac{{d(\ln S)}}{{dD}} =  - \frac{1}{{{D_0}}} = a_{MT},
\end{equation}
which concludes the proof.
\end{proof}

\begin{lemma}
$\ln b_{MT}$ is approximated by the intercept of the linear regression of $\ln S$ versus dose, computed using experimental data beyond the shoulder region of the curve.
\end{lemma}
\begin{proof}

In the deep-kill regime ($S \ll 1$, corresponding to a high dose), the exponential term of the MT survival model has very small values $e^{-D/D_0} \ll 1.$ Using the first-order approximation
\begin{equation}
\left(1 - x\right)^{b_{MT}} \approx 1 - b_{MT} x 
\quad \text{for } x \ll 1,
\end{equation}
with $x = e^{-D/D_0}$, we obtain
\begin{equation}
\left(1 - e^{-D/D_0}\right)^n 
\approx 1 - n e^{-D/D_0}.
\end{equation}

\noindent Substituting into the survival model gives
\begin{equation}
S 
= 1 - \left(1 - e^{-D/D_0}\right)^{b_{MT}}
\approx 1 - \left(1 - b_{MT} e^{-D/D_0}\right)
= b_{MT} e^{-D/D_0}.
\end{equation}

\noindent Taking logarithms yields the linear approximation
\begin{equation}
\ln S 
\approx \ln b_{MT} - \frac{D}{D_0}.
\label{Eq_linear_highD}
\end{equation}

\noindent Using the previous Lemma to obtain $D_0=\frac{1}{\alpha_{MT}}$ and setting $D=0$ in the above equation, we obtain $c=lnS_0$, where $S_0=S(D=0)$ is the intercept of the linear regression with the $logS$ axis, using only high-dose data, e.g., $S >0.1$. 

Please note that \eqref{Eq_linear_highD} does not model the shoulder of the survival curve, only the linear part; therefore, when there is a shoulder in the survival data, $S_0>1$. Thus, the number of targets per cell is obtained by $b_{MT} = e^c$~\cite{CellSurvivalCurves2024}.
\end{proof}

These analytically obtained estimates were subsequently used as starting points for full nonlinear regression of the MT model, improving numerical stability and convergence while allowing the refinement of both parameters.

\subsection{Parameterization of the LQ model}

Additionally, for the LQ model, the constants $a_{LQ}$ and $b_{LQ}$ of (5) need to be calculated for each microorganism by minimizing the error with respect to the $N$ experimental measurements $(D_i, S_i)$. By applying (5) to each experimental measurement and taking the logarithm, we can rewrite the linear equations in matrix form
\begin{equation}
\mathbf{D} \cdot \mathbf{P} = \mathbf{S}_\mathbf{L},
\end{equation}
where $\mathbf{P} = [a_{LQ}, b_{LQ}]^T$, $\mathbf{D}$ is the 2-column matrix of the doses of the points, where each row is $(D_i, D_i^2)$, and $\mathbf{S_L}$ is the column matrix of the negative logarithmic survivals:
\begin{equation}
\mathbf{D} = \begin{bmatrix}
D_1 & D_1^2 \\
D_2 & D_2^2 \\
\vdots & \vdots \\
D_N & D_N^2
\end{bmatrix}, \quad
\mathbf{S_L} =-\begin{bmatrix}
\ln S_1 \\
\ln S_2 \\
\vdots \\
\ln S_N
\end{bmatrix}.
\end{equation}
A simple linear regression is performed using the least squares method to solve this system and estimate the final values of the parameters $a$ and $b$.

\subsection{Constrained parameterization of the bi-exponential TSD model}

To ensure biologically meaningful parameter estimates in the TSD bi-exponential survival model \eqref{eq_TSDmodel} we need to enforce  $0<w<1$ and $a_{TSD}>b_{TSD}>0$, modeling two subpopulations of the microorganism: one larger with greater radio-sensitivity corresponding to the first term of the equation and a smaller, more resilient one. To this end, the parameters were reparameterized using unconstrained auxiliary variables. Specifically, the mixture weight was expressed via a sigmoid transformation,
\begin{equation}
w=\sigma(\gamma)=\frac{1}{1+e^{-\gamma}},
\label{Eq_LQ_param_w}
\end{equation}
which guaranties $0.5<w<1$ for all $\gamma\in\mathbb{R^+}$. The rate parameters were enforced to be positive and ordered by defining
\begin{equation}
b_{TSD}=e^{\eta_2}, 
\qquad
a_{TSD}=b_{TSD}+e^{\eta_\Delta},
\label{Eq_LQ_param_a_b}
\end{equation}
where $\eta_2,\eta_\Delta\in\mathbb{R}$. This formulation ensures $a_{TSD}>b_{TSD}>0$ for all real values of the auxiliary variables and avoids nonphysical solutions (e.g., negative weights or rate reversal) during nonlinear optimization, while allowing unconstrained numerical estimation employing the least squares method with the iterative Gauss-Newton algorithm.

\subsection{Confidence intervals for model parameters}

The available datasets consist of a limited number of experimental points, typically without repeated measurements. Therefore, to estimate confidence intervals for the model parameters, a parametric bootstrap approach was employed. In this framework, synthetic datasets are generated via Monte Carlo sampling from the fitted model and are repeatedly refitted, allowing for the approximation of the sampling distribution of the estimated parameters.

The experimental data correspond to UVC doses associated with microbial reductions ranging from 1-log ($10^{-1}$) to 6-log ($10^{-6}$). To account for uncertainty in the estimated inactivation parameters, an experimental relative dose uncertainty was introduced in the form of a coefficient of variation (CV). This choice reflects the predominantly multiplicative nature of radiometric uncertainties, arising from factors such as irradiance calibration, spatial non-uniformity, and geometric effects.

According to the EPA Ultraviolet Disinfection Guidance Manual (UVDGM) ~\cite{EPA_UVDGM_2006}, which serves as a primary technical reference for UV validation, the combined uncertainty associated with UV sensors, spectrophotometers, calibration drift, spectral mismatch, as well as flow inhomogeneity and lamp aging can reach up to 15\%. In this study, a representative value of $\mathrm{CV} = 0.10$ was adopted.

Since UVC dose is defined as the product of irradiance and exposure time, and the associated uncertainties are predominantly multiplicative, dose errors were modeled using a log-normal distribution. Specifically, the perturbed (noisy) dose was expressed as
\begin{equation}
D_k^{\ast} = D_k \exp(\varepsilon_k),
\qquad \varepsilon_k \sim \mathcal{N}(0,\sigma_{\ln D}^2),
\label{eq_eps_k}
\end{equation}
where $D_k$ denotes the nominal dose corresponding to a given survival level, in our case $10^{-k}$, 
which is equivalent to the natural logarithm of Dose taking values from a normal distribution, or the dose follows the lognormal distribution:
\begin{equation}
\ln D_k \sim \mathcal N\!\left(\ln \hat D_k(\boldsymbol{\theta}),\,\sigma_{\ln D}^2\right).
\label{eq:normal_logdose}
\end{equation}

For a log-normal distribution, the coefficient of variation and the log-scale standard deviation are related by
\begin{equation}
\mathrm{CV} = \sqrt{e^{\sigma_{\ln D}^2}-1},
\end{equation}
which yields
\begin{equation}
\sigma_{\ln D} = \sqrt{\ln\!\left(1+\mathrm{CV}^2\right)}.
\label{sigma_log}
\end{equation}
Thus, to generate synthetic dose measurements, the following procedure was applied for each of the four survival models:
\begin{itemize}
    \item A total of $B = 1000$ synthetic datasets were generated by independently sampling $\varepsilon_k$ according to \eqref{eq_eps_k}.
    \item For each of the $B$ perturbed datasets and for each survival level:
    \begin{itemize}
        \item Model parameters were re-estimated, yielding an empirical approximation of their sampling distributions.
        \item Confidence intervals at the $95\%$ level ($CI_{95\%}$) were computed from the 2.5th and 97.5th percentiles of the bootstrap parameter distributions.
    \end{itemize}
\end{itemize}

\begin{figure}[h]
\centering
\includegraphics[keepaspectratio,width=0.6\columnwidth]{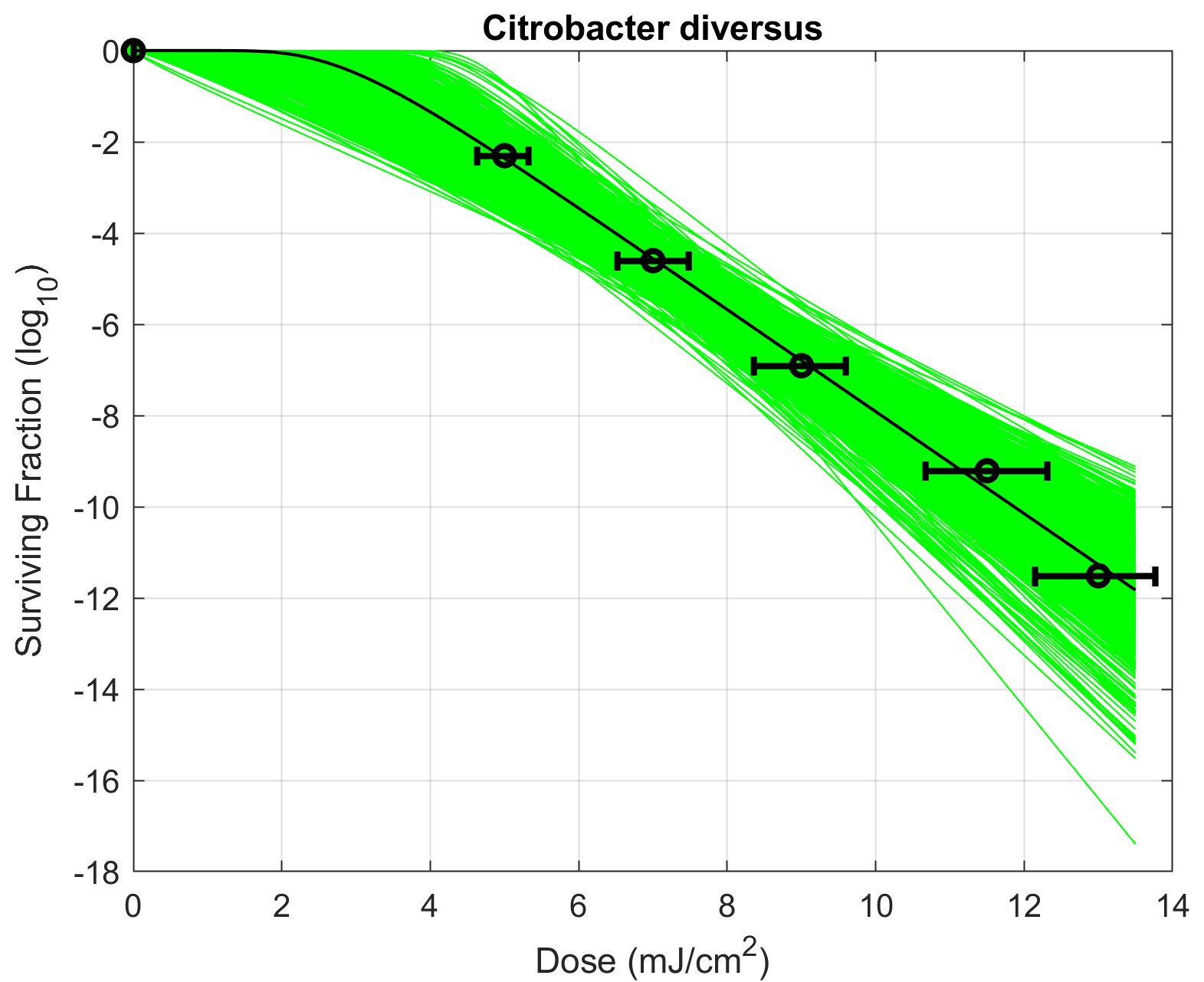}
\caption{MT survival model for \textit{Citrobacter diversus}. Black markers: measured doses; error bars: interquartile ranges from 1000 log-normal perturbations ($\mathrm{CV}=0.10$); green curves: refitted models.}
\label{Fig_conf_interval}
\end{figure}

Figure~\ref{Fig_conf_interval} illustrates the MT survival model fitted to the measured dose data (black curve and markers) for \textit{Citrobacter diversus}. A total of 1000 synthetic dose realizations were generated using a log-normal distribution with $\mathrm{CV} = 0.10$. The interquartile range is shown as horizontal error bars, while the corresponding fitted models are depicted as green curves.

Dose uncertainty was modeled as log-normal, corresponding to Gaussian noise in log-dose space with constant variance $\sigma_{\ln D}^2$. As a result, the confidence intervals are multiplicative and scale proportionally with dose, leading to an increasing absolute interval width at higher doses while maintaining constant relative uncertainty.

\section{Model comparison criteria}

Model comparison was conducted using both information-theoretic criteria and residual-based error measures. These complementary metrics provide a balanced assessment of model performance: the AIC and its AICc quantify relative model support while penalizing model complexity, whereas the MSE of the logarithm of survival (MSE$_{\ln S}$) serves as a descriptive indicator of goodness-of-fit.

\subsection{AIC}

For a model with $k$ free parameters and maximum log-likelihood $\log L(\hat{\boldsymbol{\theta}})$ as derived in the sequel, the AIC\cite{akaike2003new} is defined as
\begin{equation}
\mathrm{AIC}
=
2k - 2\log L(\hat{\boldsymbol{\theta}}).
\label{eq:AIC}
\end{equation}
AIC estimates the expected information loss incurred when a given model is used to represent the data-generating process. Smaller AIC values indicate stronger relative support among competing models fitted to the same dataset.

When the sample size $n$ is small relative to the number of parameters, AIC tends to favor overly complex models. The corrected criterion AICc \cite{hurvich1989regression} accounts for this finite-sample bias and is defined as
\begin{equation}
\mathrm{AICc}
=
\mathrm{AIC}
+
\frac{2k(k+1)}{n-k-1},
\label{eq:AICc}
\end{equation}
which is valid for $n > k+1$. In the present work, where the number of iso-survival data points is limited, AICc was used in place of AIC for all model comparisons. Differences in AICc values,
\[
\Delta_i = \mathrm{AICc}_i - \min_j \mathrm{AICc}_j,
\]
were used to assess relative model support.

\subsubsection{Iso-survival dose inversion and Log-likelihood in dose space}

Let $S(D;\boldsymbol{\theta})$ denote a parametric survival model with parameter vector $\boldsymbol{\theta}$. 
For a prescribed survival level $S_k\in(0,1)$, the model-predicted iso-survival dose $\hat D_k(\boldsymbol{\theta})$ is defined implicitly as the unique solution of
\begin{equation}
S\!\left(\hat D_k;\boldsymbol{\theta}\right)=S_k,
\label{eq:Dhat_def}
\end{equation}
i.e., $\hat D_k$ is obtained by inverting the survival curve at level $S_k$. The experimental observations are the measured doses $\{D_k\}_{k=1}^{n}$ corresponding to fixed survival levels $\{S_k\}_{k=1}^{n}$.

For the ST model, $\hat D_k$ is easily obtained in closed-form expression 
\begin{equation}
\hat D_k(a_{ST})=\frac{-\ln S_k}{a_{ST}}.
\label{eq:Dhat_exp}
\end{equation}
Solving the LQ model \eqref{Eq_LQ-model} for the dose yields the quadratic equation $\alpha_{LQ} \hat D_k+\beta_{LQ} \hat D_k^2=-\ln S_k$, whose physically admissible (nonnegative) solution is
\begin{equation}
\hat D_k(a_{LQ},b_{LQ})
=
\frac{-a_{LQ}+\sqrt{b_{LQ}^2+4b_{LQ}(-\ln S_k)}}{2b_{LQ}},
\qquad \beta>0.
\label{eq:Dhat_LQ}
\end{equation}
For the remaining two survival models (MT and TDS), analytical inversion is not available in closed form; therefore, numerical inversion is performed using the Newton--Raphson root-finding method.
For the MT model \eqref{Eq_MT-model}
$\hat D_k(D_0,n)$ is obtained numerically as the unique root of
\begin{equation}
1-\left(1-e^{-\hat D_k/D_0}\right)^{b_{MT}} - S_k = 0, \quad \hat D_k\ge 0.
\label{eq:Dhat_MT_root}
\end{equation}
Similarly, for the TDS Bi-exponential model \eqref{eq_TSDmodel}, $\hat D_k(w,a_1,a_2)$ is obtained numerically as the unique root of
\begin{equation}
w\,e^{-a_{TSD} \hat D_k}+(1-w)\,e^{-b_{TSD} \hat D_k}-S_k=0, \quad \hat D_k\ge 0.
\label{eq:Dhat_biexp_root}
\end{equation}

As described in the previous subsection, experimental uncertainty is primarily dominated by relative (multiplicative) dose errors. Accordingly, measured doses are modeled as log-normal perturbations around the model-predicted values
\begin{equation}
D_k=\hat D_k(\boldsymbol{\theta})\,\exp(\varepsilon_k),
\qquad
\varepsilon_k\sim\mathcal N(0,\sigma_{\ln D}^2),
\label{eq:lognormal_dose}
\end{equation}
which is equivalent to
\begin{equation}
\ln D_k \sim \mathcal N\!\left(\ln \hat D_k(\boldsymbol{\theta}),\,\sigma_{\ln D}^2\right).
\label{eq:normal_logdose2}
\end{equation}

The corresponding log-likelihood was formulated based on standard likelihood theory for transformed variables and nonlinear regression~\cite{CasellaBerger}. It is well established that minimizing squared residuals in a transformed space is equivalent to maximum likelihood estimation under the assumption of Gaussian errors~\cite{BatesWatts1988}. Accordingly, the log-likelihood can be expressed as follows
\begin{equation}
\log L(\boldsymbol{\theta})
=
-\sum_{k=1}^{n}\ln D_k
-n\ln\!\left(\sigma_{\ln D}\sqrt{2\pi}\right)
-\frac{1}{2\sigma_{\ln D}^2}
\sum_{k=1}^{n}
\left[\ln D_k-\ln \hat D_k(\boldsymbol{\theta})\right]^2.
\label{eq:logL_Dhat}
\end{equation}
The standard deviation $\sigma_{\ln D}$ is fixed and common across all candidate models, according to \eqref{sigma_log}. Consequently, the first two terms in Eq.~\eqref{eq:logL_Dhat} are constant with respect to $\boldsymbol{\theta}$ and therefore do not affect relative model comparisons.

\subsection{MSE of log-survival}

As a descriptive measure of goodness of fit, the MSE of the logarithm of survival was computed as
\begin{equation}
\mathrm{MSE}_{\log S}
=
\frac{1}{n}
\sum_{k=1}^{n}
\left[
\log S_k - \log \hat S(D_k;\hat{\boldsymbol{\theta}})
\right]^2,
\label{eq:MSElogS}
\end{equation}
where $S_k$ denotes the observed survival fraction at dose $D_k$, and $\hat S(D_k;\hat{\boldsymbol{\theta}})$ is the corresponding model prediction.

The MSE was computed in log-survival space, $\mathrm{MSE}_{\ln S}$, rather than in linear survival space, since survival fractions span several orders of magnitude and experimental uncertainty is predominantly multiplicative. Evaluating residuals in $\ln S$ therefore provides scale-invariant weighting across dose levels and prevents the fit from being dominated by high-survival values. Unlike AIC and AICc, $\mathrm{MSE}_{\ln S}$ does not account for model complexity and is thus reported solely as an intuitive, residual-based measure of goodness-of-fit, rather than as a criterion for model selection.

\subsection{Non-quantitative criteria for cell survival model selection}
In what follows, we examine the applicability and suitability of the four survival models under different experimental data scenarios, primarily based on their mathematical properties as well as their physical and biological plausibility. The observations and conclusions presented here should be interpreted in conjunction with the quantitative criteria described above.

\subsubsection{TSD vs.\ ST vs.\ LQ model}

The TSD model~\eqref{eq_TSDmodel} is particularly suitable when the dose--$\ln S$ curve, although monotonically decreasing, exhibits concave behavior, i.e., an initially steep decline followed by a shallower segment. 

In such cases, fitting the LQ model~\eqref{Eq_LQ-model} typically leads to a negative value of $\beta_{LQ}$ in order to reproduce the observed concavity. However, this results in a non-monotonic survival curve with a minimum at
\begin{equation}
D^* = -\frac{a_{LQ}}{2b_{LQ}},
\label{Eq_D*}
\end{equation}
beyond which survival increases with dose, which is biologically implausible.
Therefore, the LQ model loses physical validity when $b_{LQ} < 0$ and the turning point $D^*$ lie within the experimental dose range (i.e., $D^* \ll D_{\max}$). In contrast, when $b_{LQ} > 0$, the turning point satisfies $D^* < 0$, and the LQ model remains monotonically decreasing, making it suitable for modeling convex survival curves.

\begin{figure}[h]
\centering
\includegraphics[keepaspectratio,width=0.6\columnwidth]{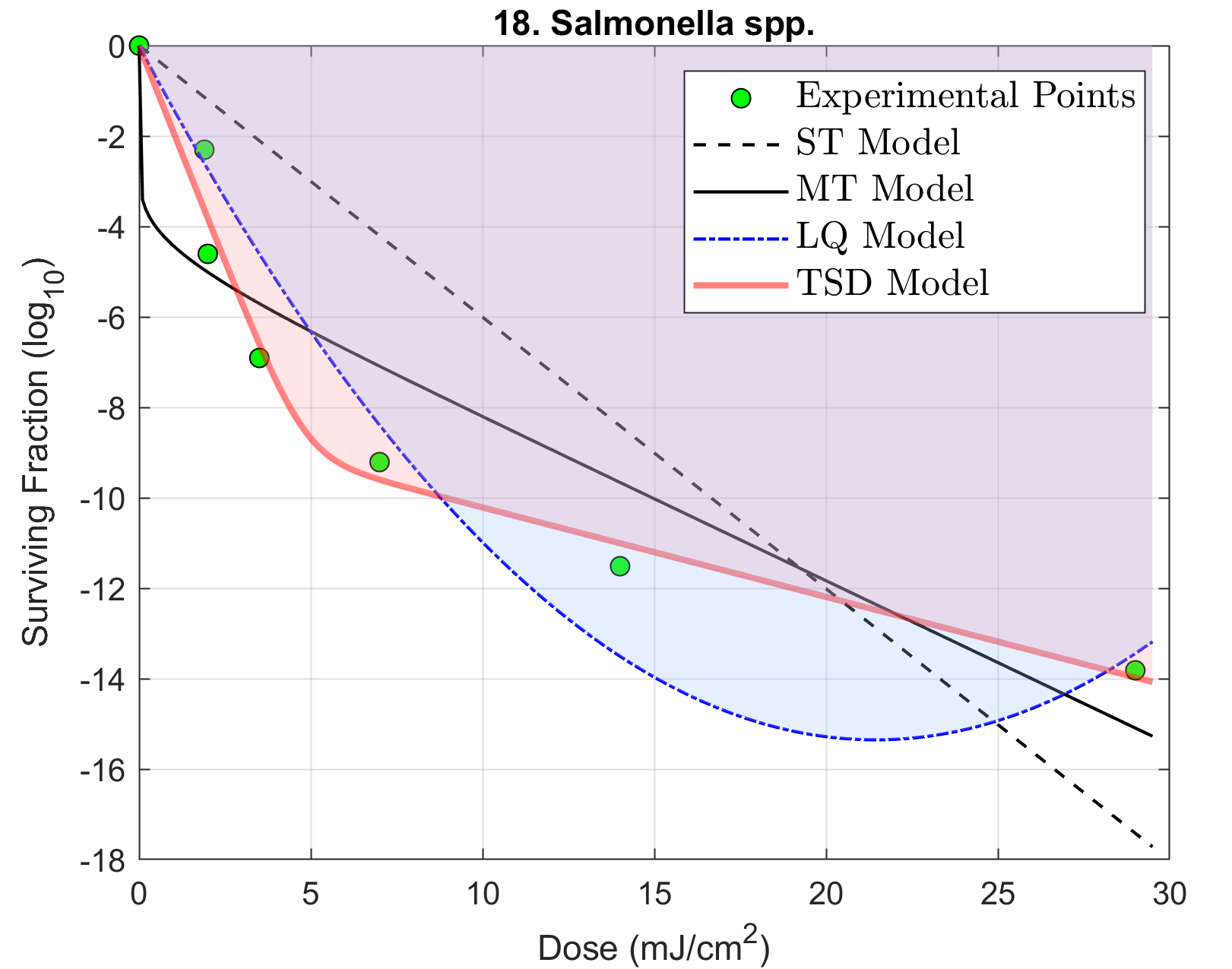}
\caption{The experimental points for \textit{Salmonella spp.} and the four parameterized survival models (more details in line 18 of Table 2).}
\label{Fig_TSD_LQ}
\end{figure}

Figure~\ref{Fig_TSD_LQ}, which compares the four survival models, clearly illustrates the aforementioned issue. The experimental data exhibit a bi-exponential trend, characterized by an initial steep linear decline in the $D$--$\log S$ domain, followed by a shallower segment. As expected, the TSD model provides the best fit according to both $\mathrm{MSE}_{\log S}$ and AICc. The LQ model ranks second in terms of $\mathrm{MSE}_{\log S}$ (2.904 vs. 13.690), while the ST model ranks second based on AICc (TSD: 53.7, ST: 457.988, LQ: 2749.161). However, the parameterization of the LQ model yields $b_{LQ} = -0.033 < 0$, resulting in a turning point at
$D^* = -\frac{a_{LQ}}{2b_{LQ}} = 21.44~\mathrm{mJ\,cm^{-2}}$, which lies well within the experimental dose range ($D^* < D_{\max} = 29~\mathrm{mJ\,cm^{-2}}$). This implies that the LQ model predicts an increase in survival beyond $D^*$, which is physically and biologically implausible. Therefore, despite its relatively favorable quantitative performance, the LQ model should be excluded as a viable alternative in this case.

In the absence of concavity, the TSD model reduces to the ST model in the limiting case $w \to 1$ and $a_1 \to a_{\mathrm{ST}}$. In such scenarios, the AICc criterion favors the ST model due to its lower model complexity. An illustrative example is shown in Fig.~\ref{Fig_MT_large_n}, where the TSD and ST models are nearly indistinguishable.

\subsubsection{MT model considerations}

The MT model~\eqref{Eq_MT-model} captures the presence of a shoulder in the $D$--$\log S$ curve, whose prominence increases with the number of hits required for inactivation, $b_{MT}$. When the shoulder is pronounced and $D_0 = \frac{1}{a_{MT}}$ is moderate or large, the estimated value of $b_{MT}$ tends to be high. In such cases, a lack of experimental data in the shoulder region (i.e., $S > 0.1$) may result in large confidence intervals for $b_{MT}$, indicating weak parameter identifiability of the MT model. Under these conditions, alternative models may be preferable, even if the MT model exhibits slightly better performance according to quantitative criteria.

Moreover, in the presence of a pronounced shoulder, the ST model—due to its inherently linear behavior in $\ln S$ and the constraint $S(0)=1$—may fail to adequately represent the data. In contrast, the LQ model can serve as a more flexible empirical alternative. In such scenarios, it is appropriate to report both the MT model and the second-best model according to AICc.

\begin{figure}[h]
\centering
\includegraphics[keepaspectratio,width=0.6\columnwidth]{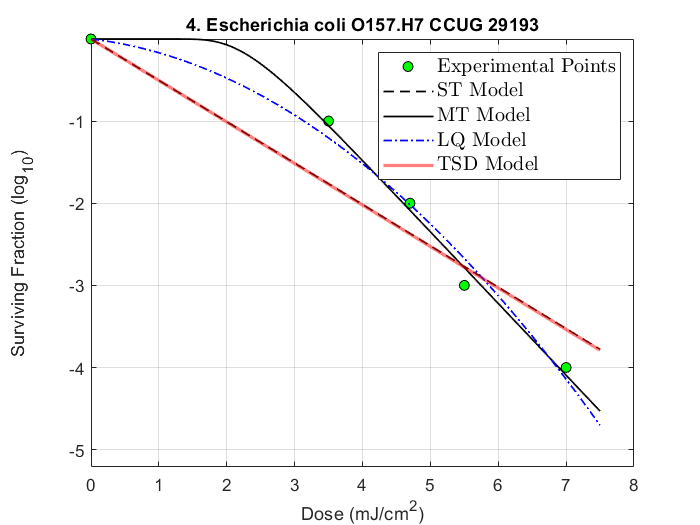}
\caption{The experimental points for \textit{Escherichia coli O157.H7 CCUG29193}, and the four parameterized survival models (more details in line 4 of Table 2).}
\label{Fig_MT_large_n}
\end{figure}

An example is shown in Fig.~\ref{Fig_MT_large_n}, where the survival data exhibit a strong shoulder but lack measurements for $S > 0.1$. The MT model provides the best fit, with the LQ model ranking second in terms of both $\mathrm{MSE}_{\log S}$ (0.351 vs. 0.904) and AICc (18.116 vs. 19.469). 
However, the estimated parameter $b_{MT} = 104.924$ is associated with a very wide 95\% confidence interval, $CI_{95\%} = [2.619,\; 13850.32]$, indicating poor identifiability. 
This also suggests that the inferred number of targets may lack clear biological interpretation. For these reasons, and in the absence of sufficient data in the shoulder region, the MT model may be a more reliable alternative for this microorganism. Accordingly, both the best-performing and the second-best models are reported in the Results section to provide flexibility in model selection.

\section{Results}
 The proposed methodology was applied to experimental dose--response data for 32 microorganisms collected from the literature, as shown in the first column of Table \ref{Table1}. The same table also summarizes, where available, key experimental details, including UV wavelength, maximum dose per unit surface, exposure time, distance from the source, and maximum log-reduction. Due to inconsistencies in reporting across studies, these parameters are not uniformly available and are therefore not emphasized in the analysis. We acknowledge the inherent heterogeneity of the compiled experimental data, as well as the potential influence of environmental conditions and the microbial physiological state (e.g., cell cycle phase) at the time of irradiation on the measured survival response~\cite{McMahon2018}.

Comprehensive results for all 32 microorganisms are presented in Table~2. The second column reports the maximum dose $D_{\max}$ for each microorganism, followed by columns containing $\mathrm{MSE}_{\log S}$ and AICc values. The notation ``N/A'' indicates cases where AICc could not be computed due to an insufficient number of experimental data points. Subsequent columns provide the estimated parameters for all four models. For the ST model, only the parameter $a_{ST}$ is reported. For the MT and LQ models, the parameters $(a, b)$ correspond to $(a_{MT}, b_{MT})$ and $(a_{LQ}, b_{LQ})$, respectively. For the TSD model, all three parameters $(a_{TSD}, b_{TSD}, w)$ are included. The "Remarks" column contains concise but informative comments regarding model performance and parameter interpretation. The rightmost column presents the experimental data alongside the fitted curves for all four models. Model selection for each microorganism was primarily based on the MSE in log-survival space, $\mathrm{MSE}_{\log S}$, computed between the logarithm of the measured survival and the corresponding model predictions. The model achieving the minimum $\mathrm{MSE}_{\log S}$ was selected as the most appropriate, with AICc and qualitative criteria used as complementary indicators.

\begin{figure*}[h]
\centering
\subfigure[]{\includegraphics[width=0.45\columnwidth]{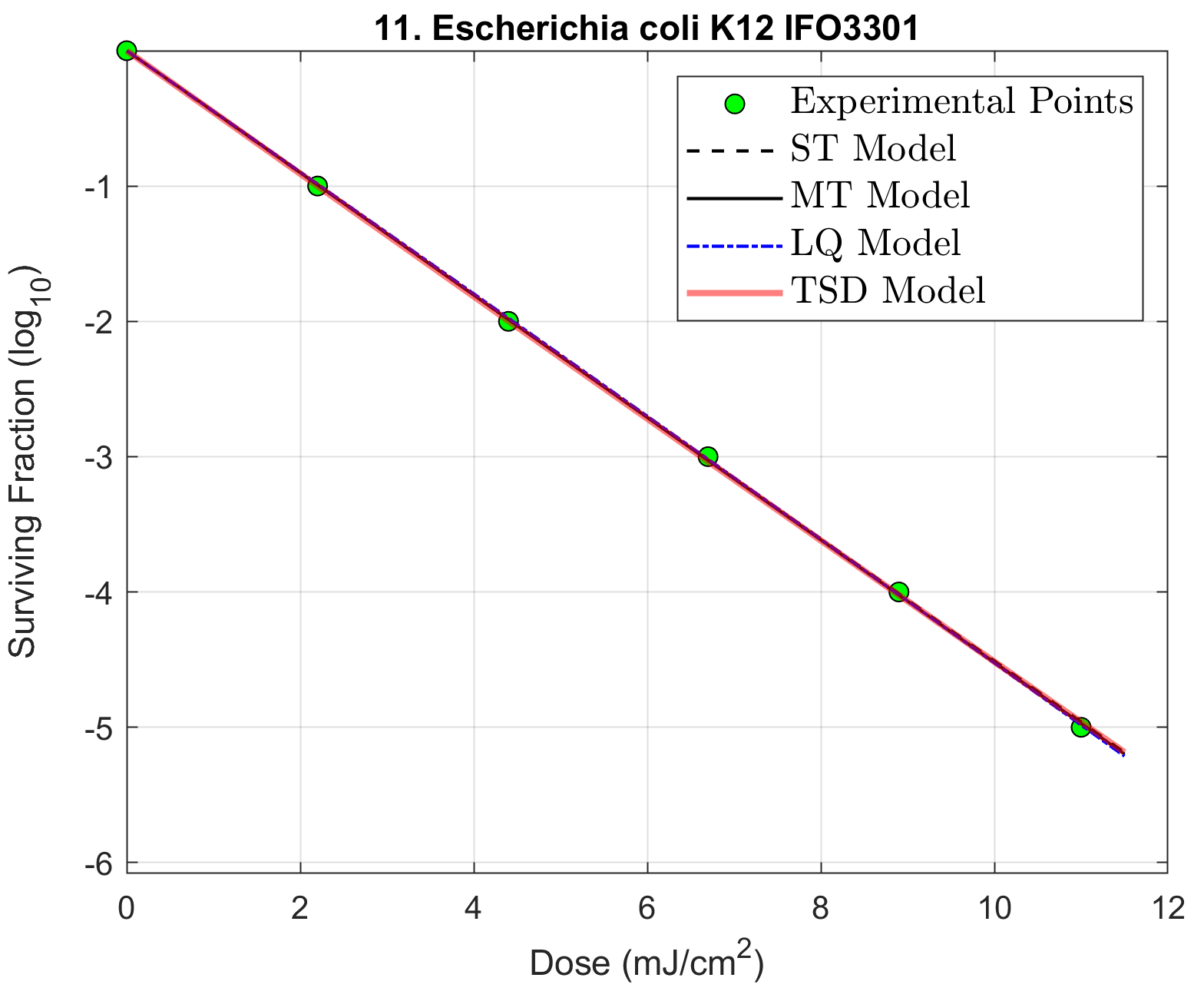}}
\subfigure[]{\includegraphics[width=0.45\columnwidth]{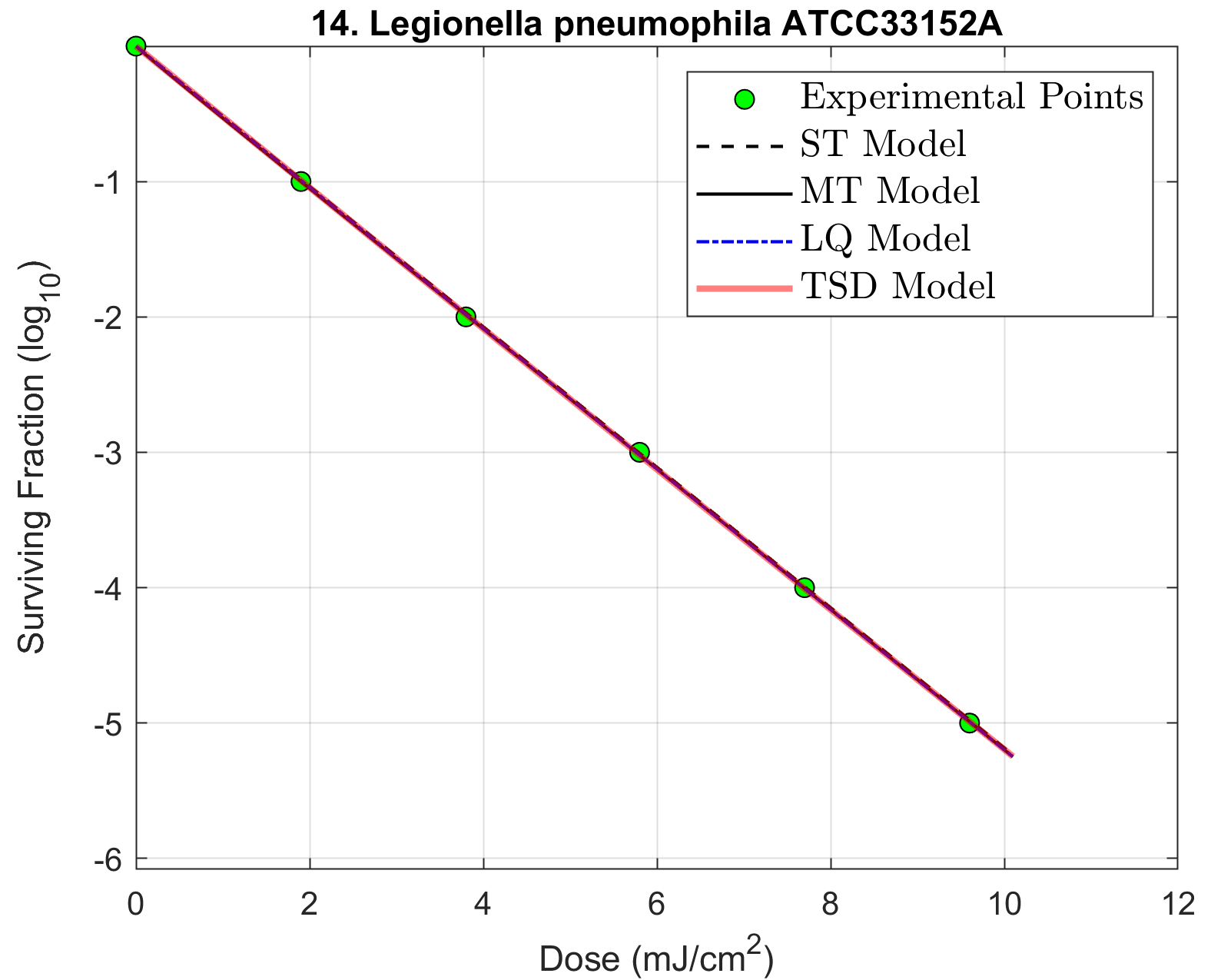}}
\caption{Examples of microorganism UVC survival curves, where ST is optimal (AICc); other models mimic it, (a) \textit{Escherichia coli K12 IFO3301} and (b) \textit{Legionella pnumophila ATCC33152A} (lines 11 and 14 in Table 2).}
\label{Fig_ST_2graphs}
\end{figure*}

The resulting regressions can be broadly categorized into distinct groups. A subset of microorganisms (lines~3, 7, 11, 14, 16, 19, 25, 26, and 27 in Table~2) exhibits an approximately linear relationship between $\log S$ and dose $D$. In these cases, the ST model typically achieves the lowest or near-lowest $\mathrm{MSE}_{\ln S}$ and the best AICc due to its single-parameter structure. The MT and LQ models effectively reduce to the ST form, with $b_{MT} \approx 1$ and $\beta_{LQ} \approx 0^{+}$, respectively. Similarly, the TSD model collapses to the ST model when $w \approx 1$ (i.e., $1-w \approx 0$) and $a_{TSD} \approx a_{ST}$, rendering $b_{TSD}$ irrelevant. In addition, the TSD model may exhibit non-identifiability, as multiple combinations of $(w, a_{TSD}, b_{TSD})$ can produce equivalent linear behavior. Consequently, when the ST model achieves the lowest AICc and a very low $\mathrm{MSE}_{\ln S}$, it is selected as the most appropriate model for the corresponding microorganism. Representative examples of such survival curves are shown in Fig.~\ref{Fig_ST_2graphs}.

\begin{figure*}[h]
\centering
\subfigure[]{\includegraphics[width=0.47\columnwidth]{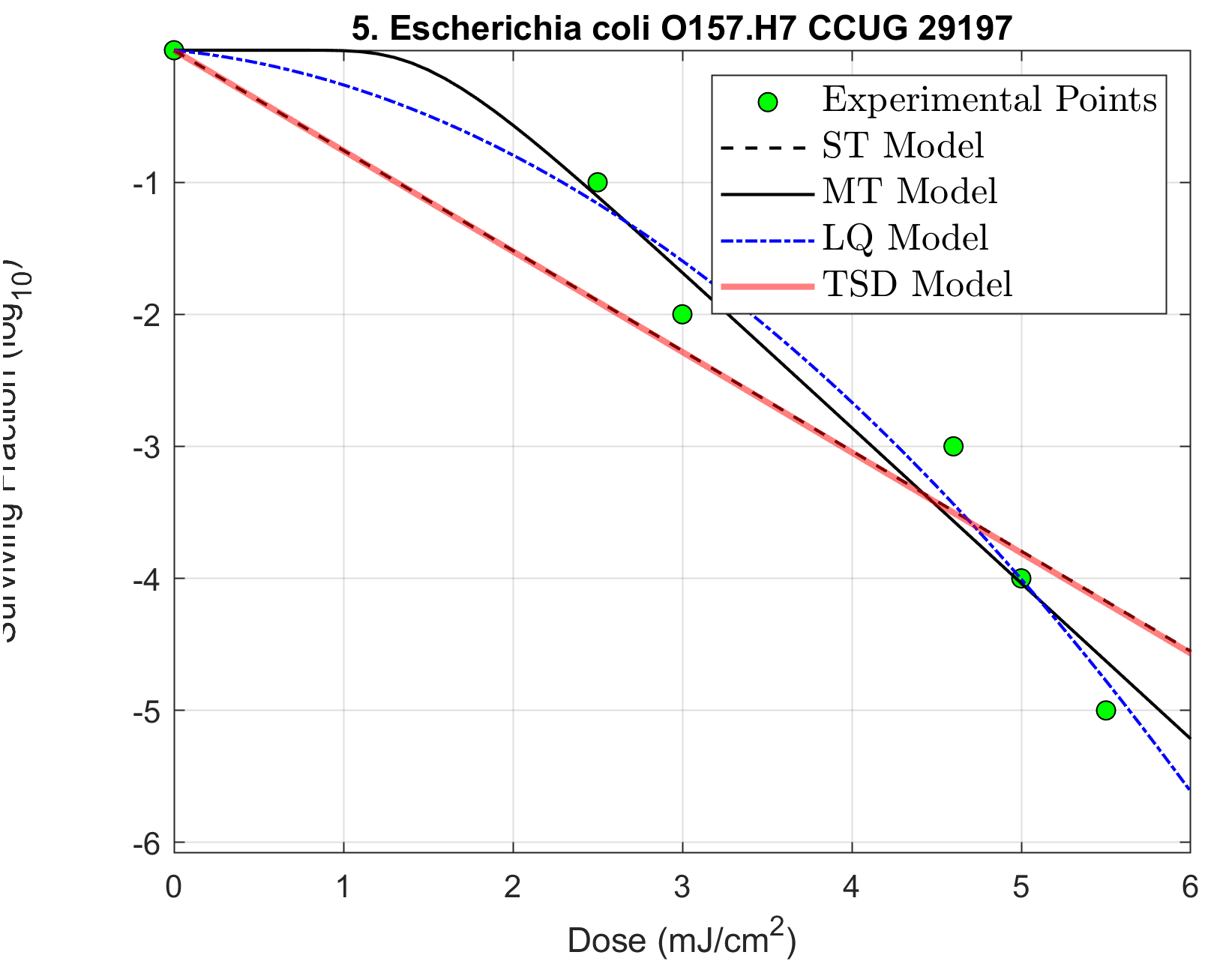}}
\subfigure[]{\includegraphics[width=0.45\columnwidth]{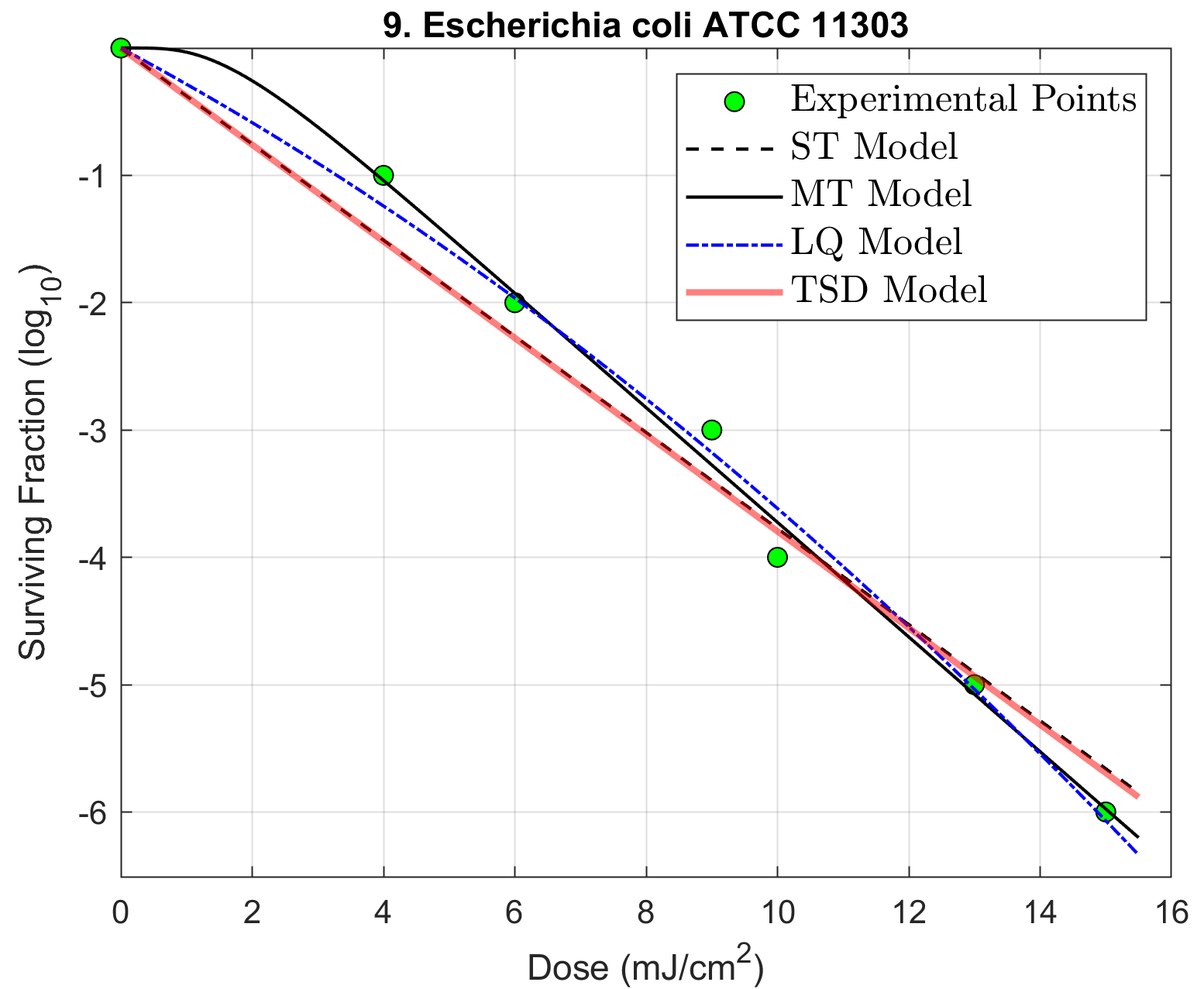}}
\caption{Examples of dose-response curves with (a) strong shoulder \textit{E. coli O157.H7 CCUG 29197} (line 5 in Table 2) and (b) mild shoulder \textit{E. coli ATCC 11303} (line 9 in Table 2); MT is weakly identifiable in (a) but reliable in (b).}
\label{Fig_shoulder_2graphs}
\end{figure*}

A subset of microorganisms exhibits a mild shoulder at low doses, followed by an approximately linear decay at higher doses. In such cases, the MT model typically achieves the best performance according to quantitative metrics and remains identifiable, as indicated by reasonably narrow confidence intervals for $b_{MT}$. Accordingly, the MT model is selected as the most appropriate, with the LQ model often providing the second-best fit. Representative cases include microorganisms Nos.~1, 9, 15, 20, 22, and 28 in Table~2, with a typical example shown in Fig.~\ref{Fig_shoulder_2graphs}(b).

In contrast, several microorganisms exhibit a more pronounced shoulder at low doses. When experimental data in this region are sparse, the MT model becomes weakly identifiable, as reflected by very wide confidence intervals for $b_{MT}$, despite achieving favorable quantitative metrics. In such cases, the LQ model may be considered a more stable and practically useful alternative. This behavior is observed for microorganisms (lines~2, 4, 5, 6, 8, 10, 13, 17, and 24 in Table~2), with a representative example shown in Fig.~\ref{Fig_shoulder_2graphs}(a).

\begin{figure*}[h]
\centering
\subfigure[]{\includegraphics[width=0.43\columnwidth]{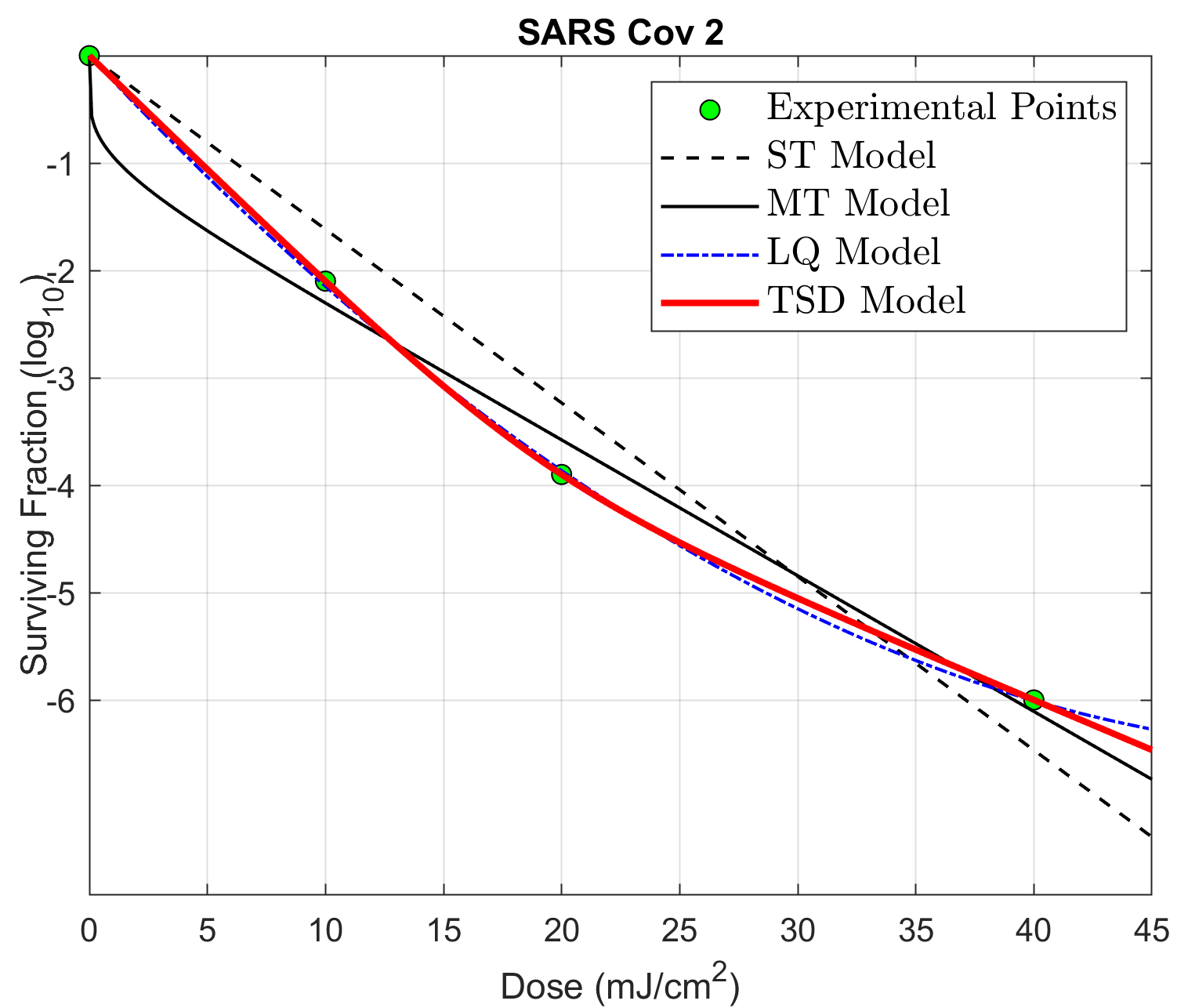}}
\subfigure[]{\includegraphics[width=0.45\columnwidth]{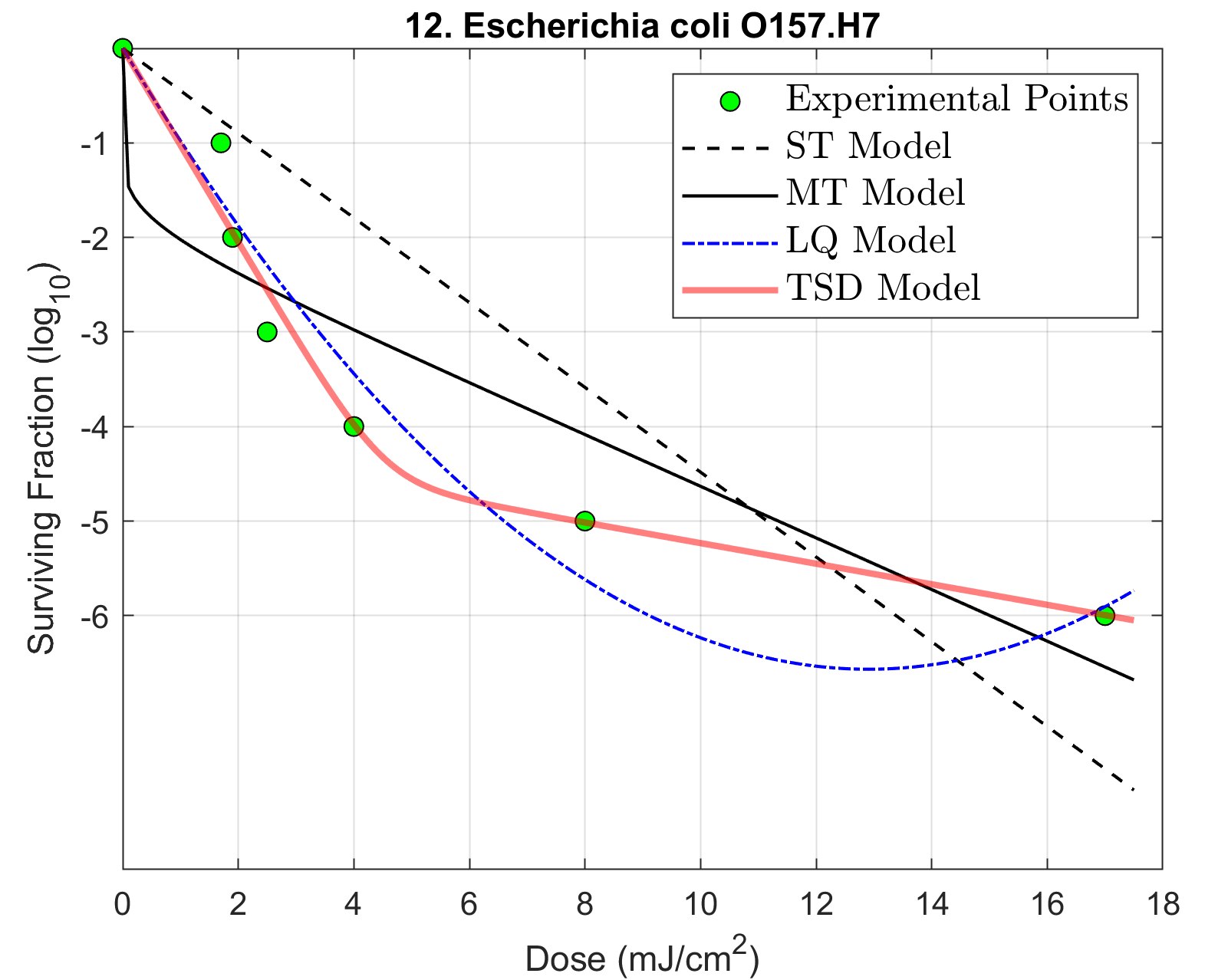}}
\caption{ Examples of concave dose-survival curves for (a) \textit{SARS-CoV-2}  (last line in Tabe 2) and (b) \textit{E. coli O157:H7} (line 12 in Table 2); TSD provides the best fit. LQ in sompetitive in (a) since $D^*>>D_{max}$, but not plausible in (b) ($D^*<D_{max}.$)}
\label{Fig_TSD_2graphs}
\end{figure*}

A number of microorganisms exhibit TSD behavior, characterized by an initial steep decline followed by a slower decay phase (e.g., lines~12, 18, 21, 23, 29, 30, and 32 in Table~2). In these cases, the TSD model is the most appropriate, as the remaining models cannot adequately reproduce this behavior. Typical examples are presented in Fig.~\ref{Fig_TSD_2graphs}. Although the LQ model may achieve competitive quantitative performance for such concave survival curves, its validity must be carefully assessed. In particular, the turning point $D^*$ (defined in~\eqref{Eq_D*}) should satisfy $D^* \gg D_{\max}$ or $D^* < 0$ to ensure physical plausibility. For example, in the case of \textit{SARS-CoV-2} (Fig.~\ref{Fig_TSD_2graphs}), both the TSD and LQ models provide comparable fits, and additional experimental data are required to clearly distinguish between them.

Finally, as summarized in Table~2, the ST model was selected as the most appropriate for 9 microorganisms, the MT model for 5, the LQ model for 10, and the TSD model for 8 out of the 32 microorganisms considered in this study.

\section{Discussion and conclusions}

The aim of this work was to develop a unified mathematical and statistical framework for analyzing microbial survival under ultraviolet irradiation using four widely adopted dose--response models (ST, MT, LQ, and TSD). The analysis focused on their structural properties, parameter identifiability, and biologically meaningful parameterizations. A parametric bootstrap approach was introduced to estimate confidence intervals in the absence of replicate measurements, and objective model selection criteria were established based on AICc, goodness-of-fit, and parameter interpretability. The framework was applied to UVC dose--response data for 32 microorganisms, enabling consistent parameter estimation and systematic selection of the most appropriate model for each case.

Several important conclusions can be drawn from both the theoretical analysis and the empirical results:
\begin{itemize}
    \item When the $(D, \log S)$ relationship is linear, the MT, LQ, and TSD models effectively reduce to the ST model. Due to its simplicity, the ST model is preferred, typically achieving the lowest AICc. In such cases, the TSD model may become non-identifiable, as multiple parameter combinations can yield equivalent behavior.
    
    \item When the survival curve exhibits a shoulder, the MT model is theoretically the most appropriate, followed by the LQ model. However, in the absence of experimental data in the shoulder region (i.e., $S > 0.1$), the parameter $b_{MT}$ often becomes weakly identifiable, resulting in large estimates and wide confidence intervals, despite favorable quantitative metrics.
    
    \item If experimental data are available in the shoulder region, or if the shoulder is not pronounced, the identifiability of $b_{MT}$ improves significantly. In such cases, the MT model can be reliably selected.
    
    \item The LQ model may serve as a competitive alternative in shoulder-type behavior, as it can reproduce the observed curvature without suffering from identifiability issues, although it typically exhibits inferior AICc and $\mathrm{MSE}_{\log S}$ values. The TSD model is generally not suitable in these cases, as it is designed to capture concave behavior. Similarly, the ST model may be inadequate due to its inherently linear form.
    
    \item For concave $\log S$ curves, characterized by an initial steep decline followed by a shallower phase, the TSD model is typically the only suitable choice, as the other models cannot reproduce this behavior. The proposed reparameterization ensures parameter identifiability in these cases. Although the LQ model can approximate concavity, its validity requires that the turning point $D^*$ satisfies $D^* \gg D_{\max}$.
    
    \item The proposed parametric bootstrap methodology, based on Monte Carlo sampling with log-normal perturbations, provides an effective approach for uncertainty quantification in the absence of replicate measurements, which is a common limitation in the available experimental datasets.
\end{itemize}

It should be noted that the survival datasets analyzed in this study were obtained under heterogeneous experimental conditions. Key factors such as irradiation wavelength, geometry, surface properties, source characteristics, environmental conditions, and biological variability are not explicitly included in current survival models but are instead implicitly reflected in the fitted parameters through radiometric dose measurements. However, these factors are rarely reported in sufficient detail to be modeled directly.

The proposed methodology partially addresses this variability by incorporating measurement uncertainty through parametric bootstrapping and the estimation of confidence intervals, thereby providing uncertainty bounds on the fitted parameters. Nevertheless, the aim of this work is not to derive universal, condition-independent parameters but to investigate the properties of different survival models, establish model selection criteria, and identify, for each microorganism, the model that best represents the available data.

Despite the heterogeneity of the datasets, the resulting parameterizations show a notable degree of consistency, supporting the robustness of the proposed framework. The models should not be interpreted as stand-alone predictors of disinfection efficacy in complex environments; rather, they are intended for dose–response modeling under specific irradiation conditions.

Looking ahead, such models can be combined with light propagation simulators to enable spatially resolved assessments of UVC disinfection in realistic settings. For example, Monte Carlo–based tools can model photon transport, accounting for reflection, scattering, and absorption \cite{Baltadourou2021}. As more detailed and standardized experimental data become available, such integrated approaches are expected to provide more accurate and practically relevant predictions.

\bibliography{References}

\includepdf[
    pages=-,
    pagecommand={},
    fitpaper=true
]{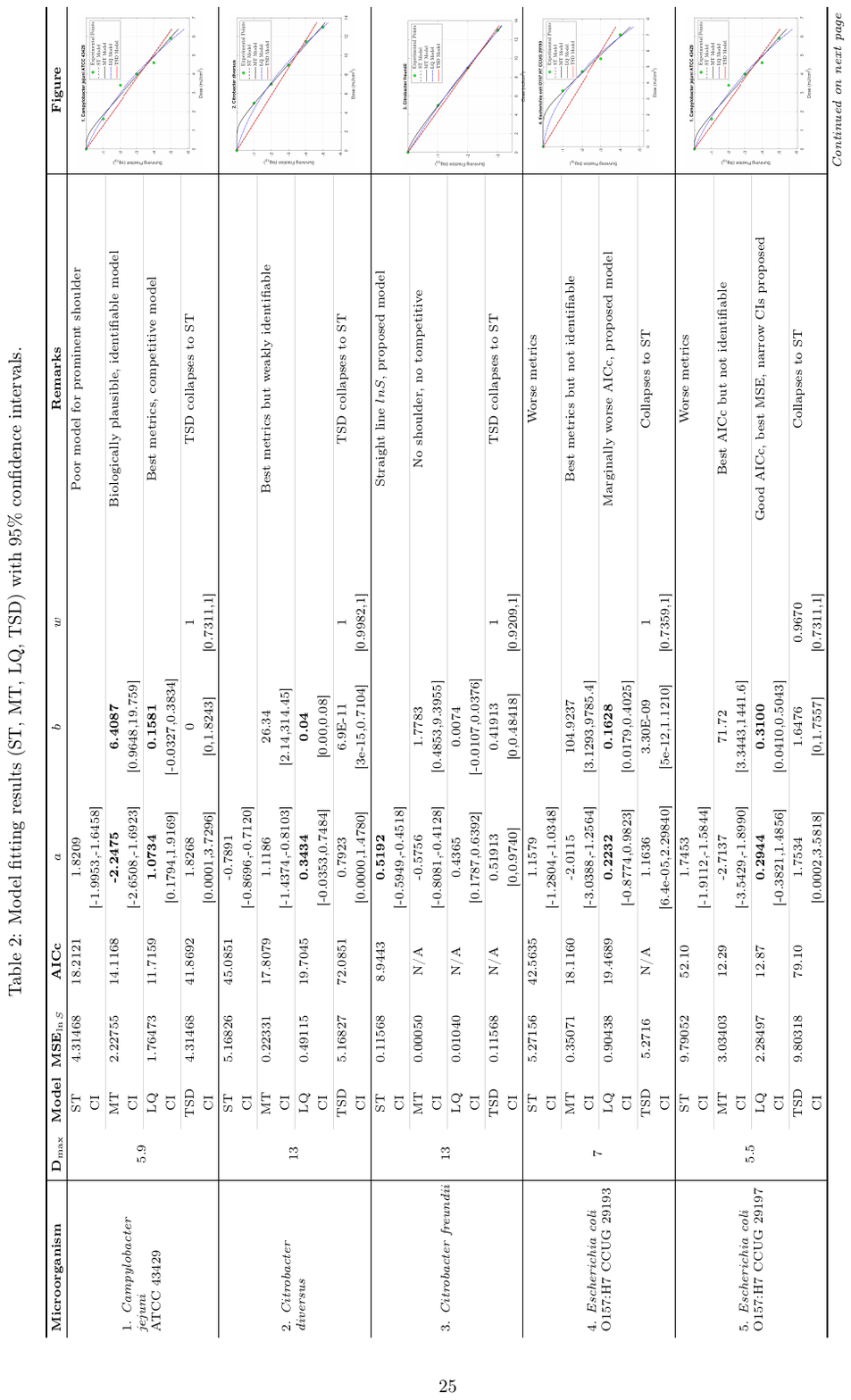}
\end{document}